\documentclass[12pt]{article}
\usepackage{amsmath,amsfonts,amssymb,latexsym}
\usepackage{cite}
\usepackage{xcolor}

\def\be{\begin{equation}}
\def\ee{\end{equation}}
\def\bq{\begin{eqnarray}}
\def\eq{\end{eqnarray}}
\def\beq{\begin{eqnarray*}}
\def\eeq{\end{eqnarray*}}
\numberwithin{equation}{section}

\begin{document}

\title{\huge{Quasi-isotropic Asymptotic Expansions in Varying Speed of Light Cosmologies}}
\author{\large{\textsc{Dimitrios Trachilis}\thanks{Dimitrios.Trachilis@aum.edu.kw}}\\
College of Engineering and Technology, \\
American University of the Middle East, Egaila 54200, Kuwait}
\date{}
\maketitle
\newpage
\begin{abstract}
\noindent  We generalize the quasi-isotropic solution of the Einstein equations near a cosmological initial singularity to models with a varying speed of light and/or a varying gravitational `constant'.
We construct a formal asymptotic series expansion in a synchronous reference system, taking into account the additional degrees of freedom introduced by the two varying `constants'.
By applying the Landau-Lifshitz quasi-isotropic method, we go beyond the standard results of previous studies that based on the Friedmann-Lema\^{i}tre-Robertson-Walker (FLRW) model, providing a more general asymptotic analysis of varying speed of light cosmologies.
We show that the resulting solutions do not contain the required number of arbitrary functions to qualify as general solutions of these theories.\\

\end{abstract}

\newpage
\tableofcontents
\newpage
\section{Introduction}

Varying speed of light (VSL) theories were proposed by Moffat \cite{1,2} as alternatives to inflationary models \cite{3}, in order to study the features of the cosmological horizon and the flatness puzzle.
It was proposed that the speed of light could have been significantly larger during the early Universe, allowing a causal connection to all regions, and after a critical time it was fixed at its present value.
In this formulation, with the obvious violation of the Lorentz invariance, changing the speed of light $c$ in the early Universe was an alternative to the change in the matter content of the Universe \cite{4,5,6}.
Although standard VSL cosmological models rely on solutions based on the behavior of isotropic and homogeneous (FLRW) metric, a deeper understanding requires the study of possible inhomogeneous solutions near the initial singularity.
In a spatially homogeneous and isotropic Universe, the Friedmann equations are still valid under the assumption of possible time variations in the speed of light $c=c(t)$ and the gravitational `constant' $G=G(t)$ (see \cite{7}).
In \cite{8}, transforming the Friedmann equations with time-varying $c$ and $G$ into a dynamical system enabled a detailed behavioral analysis using the method of asymptotic splittings \cite{9}.
In fact, the assumption that both $c$ and $G$ are proportional to the expansion rate of the Universe provided some interesting results about the form of the scale factor $a=a(t)$ which are consistent with the results in \cite{10}.
The same assumption for the speed of light was employed in \cite{11} in order to achieve a reduction of the dynamics of VSL models.
Two procedures were proposed for the study of the evolution of VSL theory.
In particular, one can find an analysis based firstly on the reduction of the dynamics of VSL models to a two-dimensional Hamiltonian dynamical system, and secondly on the parameterization of the time variable (see also \cite{12}).
The quasi-isotropic expansion of the metric tensor near the initially singularity, presented in \cite{13} and \cite{14}, was an attempt to find inhomogeneous cosmological asymptotic solutions of the Einstein equations in the case of vacuum or a radiation-dominated Universe (see also \cite{15} and \cite{16}).
This approximation technique was also used to investigate the form of the solution near the initial or sudden singularity and determine the genericity of the solution in modified gravity theories \cite{17,18,19} and Brans-Dickey theory \cite{20}.
In this paper, we apply the quasi-isotropic approximation of the metric in the synchronous system near the initial singularity in the context of VSL cosmologies.
According to the Albrecht-Magueijo model \cite{5}, it is possible to describe VSL cosmologies using the known Einstein equations $R_{ij}-\frac{1}{2}g_{ij}R-\Lambda g_{ij}=\frac{8\pi G}{c^4}T_{ij}$, but this requires a specific choice of the time coordinate as the comoving proper time.
Using this specific time as the comoving proper time, the Einstein tensor $G_{ij}=R_{ij}-\frac{1}{2}g_{ij}R$ is the same as in the standard model, namely its covariant divergence vanishes.
However, in a varying $c$ and/or $G$ model the vanishing of the covariant derivative of the Einstein tensor does not simply imply the vanishing of the stress-energy tensor $T_{ij}$ on the right-hand side of the Einstein equations.
Therefore, energy and momentum will not be conserved as usual.
In particular, instead of $T_{ij}$, the expression that has to vanish is the covariant divergence of $\frac{G(t)}{c(t)}T_{ij}$.
Based on previous homogeneous and isotropic solutions, this paper allows us to determine whether the simplest choice of formal quasi-isotropic inhomogeneous series expansion of the metric provides a solution that can qualify as a general one near the initial singularity.
Accordingly, we explore the impact of the varying fundamental `constants' $c$ and $G$ on the final balance of this theory, and we generalize the results found for homogeneous and isotropic models to the more general model used in this work.

The remainder of this paper is organized as follows: In Section 2, we set up the 3+1 split gravitational field equations with varying $c$ and $G$ under a synchronous-type framework and discuss the counting of functional degrees of freedom. In Sections 2.1--2.5, we implement the Landau-Lifshitz quasi-isotropic method to derive the formal asymptotic series expansions for the spatial metric, energy density, and fluid velocity near the initial singularity. Section 2.6 maps out the exact parameter domains ensuring series consistency and analyzes the physical thresholds of specific fluid indices. Finally, Section 3 presents our discussions and conclusions, summarizing our core findings and outlining prospective avenues for future research.

\section{Solution with varying $c$ and $G$}

It should be noted that the formulation of the metric in a synchronous reference system is not merely a mathematical convenience but a fundamental aspect of our VSL approach. We adopt the ``structural prescription'' framework \cite{4}, where the variation of the speed of light $c(t)$ and the gravitational coupling $G(t)$ is explicitly defined with respect to a preferred cosmological rest frame. As detailed by Magueijo \cite{4}, such preferred-frame models explicitly break Lorentz invariance, rendering the synchronous time coordinate the physical parameter that dictates the evolution of the fundamental ``constants''. Consequently, time differentiation does not apply to $c$ during the computation of the Christoffel symbols, naturally yielding $\Gamma^0_{00} = 0$. Alternatively, this setup can be interpreted as working directly in a unit-lapse cosmic time coordinate system by introducing a reparameterized cosmic time $\tau$ via $d\tau = c(t) dt$, which automatically yields $g_{00}=1$ and $\Gamma^0_{00}=0$.

In this Section, we present the detailed steps by which we construct a solution of the VSL field equations near the initial singularity, in order to determine whether it contains the required number of arbitrary functions to qualify as a general solution of the system.
This is done in three steps. First, we count the number of arbitrary functions present in a general solution of the field equations.
Then, we provide all necessary metric and curvature expansions valid in a neighborhood of the initial singularity, and finally, we count the arbitrary functions present in the so-constructed solution.

\subsection{Arbitrary data in the general solution}
We start with the Einstein field equations with cosmological constant $\Lambda=0$ in the standard form  (Latin indices are for spacetime
components, Greek for spatial ones),
\be
G^i_j \equiv R^i_j - \frac{1}{2}\delta^i_jR = \dfrac{8\pi G}{c^4} T^i_j.
\label{Einstein}
\ee
The stress-energy tensor is that of a perfect fluid,
\begin{equation}
T_{j}^{i}=  (\rho +\dfrac{p}{c^2})u^{i}u_{j}-p\delta _{j}^{i},
\label{stress-energy tensor}
\end{equation}%
where $\rho c^2$ and $p$ are the energy and pressure densities of the fluid, respectively, that obey an equation of state
\be
p=(\gamma - 1)\rho c^{2}.
\label{eqstate}
\ee
For the $\gamma$-parameter, we consider that $0<\gamma \leq 2$. The unit 4-velocity $u^{i}=(u^{0},u^{\alpha })$ with $u^{0}=u_{0}$, and $u_{i}u^{i}=1$, satisfies
\be
u_{0}^{2}=1+u_\alpha u^\alpha,
\label{velocities}
\ee
which provides the information that the three arbitrary components $u^\alpha$ of the velocity vector field determine $u^0$.
In the synchronous system of spacetime coordinates, the metric takes the form,
\begin{equation}
ds^{2}=c^{2}dt^{2}-\gamma_{\alpha \beta}dx^{\alpha}dx^{\beta},
\label{met}
\end{equation}%
and the field equations (\ref{Einstein}) split into the constraint equations,
\begin{equation}
R^0_0 = \dfrac{8\pi G}{c^4} (T^0_0 - \dfrac{1}{2}T),
\label{field00}
\end{equation}
and,
\begin{equation}
R^0_\alpha = \dfrac{8\pi G}{c^4} T^0_\alpha,
\label{field0a}
\end{equation}%
and the `evolution' equations,
\begin{equation}
R^\beta_\alpha = \dfrac{8\pi G}{c^4} (T^\beta_\alpha - \dfrac{1}{2}\delta^\beta_\alpha T).
\label{fieldab}
\end{equation}

Equations (\ref{field00}), (\ref{field0a}), and (\ref{fieldab}) represent the 3+1 splitting of the generalized VSL field equations in a synchronous-type frame. Specifically, Eq. (\ref{field00}) governs the dynamic evolution of the trace of the extrinsic curvature and couples the primary kinetic driving terms of the metric directly to the total energy density ($\rho$). Eq. (\ref{field0a}) serves as the momentum constraint, linking the spatial gradients to the fluid’s inhomogeneous velocity components ($u_\alpha$). Finally, Eq. (\ref{fieldab}) dictates the spatial metric propagation driven by the spatial Ricci curvature and pressure. Collectively, they form the algebraic system into which our formal power-law series are substituted to solve for the subleading coefficients.

Counting the free data, this system of equations contains $6$ arbitrary functions that correspond to the $g_{\alpha \beta}$'s, plus $6$ corresponding to the $\dot{g}_{_{\alpha\beta }}$'s, as well as $3$ free velocity components $u_{\alpha },$ plus $1$ related to the density $\rho$.
Therefore, the system gives a total of $16$ functions. However, not all are independent due to the $4$ constraints of the system, and the $4$ general coordinate diffeomorphisms.
Thus, the number of arbitrary functions of the system that are expected in the general solution of the dynamical system is $8$.

\subsection{Generalized conservation equation}
Barrow in \cite{7} pointed out that modifications of the speed of light in the local Lorentzian frames associated with cosmological expansion are a special relativistic effect.
The lack of covariance of the resulting theory requires a specific choice of time coordinate.
In order to keep the usual form of the Friedmann equations, we choose that specific time to be comoving proper time, assuming the Universe is spatially homogeneous and isotropic.
Therefore, the speed of light $c$ and the gravitational `constant' $G$ have only time variations and the scale factor $a=a(t)$ obeys the following Friedmann equations with varying $c(t)$ and $G(t)$,
\begin{eqnarray}
(\dfrac{\dot{a}}{a})^{2} &=& -\dfrac{kc^2}{a^2}+\dfrac{8\pi G}{3}\rho ,
\label{Friedman1}\\
\dfrac{\ddot{a}}{a} &=& -\dfrac{4\pi G}{3}(\rho + \dfrac{3p}{c^2}) = -\dfrac{4\pi G}{3}(3\gamma -2)\rho ,
\label{Friedman2}
\end{eqnarray}
where the equation of state has the form (\ref{eqstate}), and $k$ is the metric curvature parameter normalized to $k=0, +1$ or $-1$, for the flat, closed, or open space, respectively.
The generalized conservation equation, including the time variations in $c$ and $G$, is obtained by differentiating eq.
(\ref{Friedman1}) with respect to time and substituting in eq. (\ref{Friedman2}).
As a result, we find,
\be
\dot{\rho}+3 \dfrac{\dot{a}}{a} (\rho + \dfrac{p}{c^2}) = -\rho \dfrac{\dot{G}}{G} + 3k\dfrac{c\dot{c}}{4\pi G a^2},
\label{conlaw1}
\ee
that is,
\be
\dot{\rho}+3\gamma \rho \dfrac{\dot{a}}{a} = -\rho \dfrac{\dot{G}}{G} + 3k\dfrac{c\dot{c}}{4\pi G a^2}.
\label{conlaw}
\ee

The local non-conservation laws for the inhomogeneous perfect fluid are expressed by $\nabla_i \left( \frac{G(t)}{c(t)^4} T^i_j \right) = 0$. Projecting this relation parallel and perpendicular to the 4-velocity field yields the generalized continuity and Euler equations, respectively. Specifically, the $j=0$ component gives $\partial_t \left( \frac{G}{c^4} T^0_0 \right) + \partial_\alpha \left( \frac{G}{c^4} T^\alpha_0 \right) + \frac{G}{c^4} \left( \Gamma^i_{ki} T^k_0 - \Gamma^k_{i0} T^i_k \right) = 0$, which reduces to $\dot{\rho} + 3\gamma \rho \frac{\dot{a}}{a} + \rho \frac{\dot{G}}{G} + \frac{1}{a^3} \partial_\alpha \left( a^3 \gamma \rho u^\alpha \right) = 0$. For the $j=\alpha$ component, we obtain $\partial_t \left( \frac{G}{c^4} T^0_\alpha \right) + \nabla_\beta \left( \frac{G}{c^4} T^\beta_\alpha \right) + \frac{1}{2} K \left( \frac{G}{c^4} T^0_\alpha \right) = 0$. Since the Einstein tensor satisfies the contracted Bianchi identities $\nabla_i G^i_j \equiv 0$, any asymptotic series constructed to solve the field equations order-by-order automatically satisfies these generalized non-conservation laws.

\subsection{Expansions}
In \cite{8}, using the method of asymptotic splittings introduced in \cite{9}, under the assumption that both functions $c=c(t)$ and $G=G(t)$ have power-law variations of the form
\be
c=c_{0}a^{n}, \quad G=G_{0}a^{m}, \quad n,m\in \mathbb {R},
\label{cGseries}
\ee
it was demonstrated that the Friedmann isotropic solution near the singularity, taken at $t=0$, which obeys an equation of state of the form (\ref{eqstate}), behaves as
\be
a \sim t^{2/3\gamma}.
\label{abeh}
\ee

Therefore, the general form of the spatial part of the metric (\ref{met}) considered as a solution of the field equations, with varying $c$ and $G$, and constructed according to the Cauchy problem will have an expansion of the form
\be
\gamma _{\alpha \beta } = a_{\alpha \beta }t^{A}+b_{\alpha \beta}t^{B}+\cdots, \quad B>A,
\label{seriesgab}
\ee
where
\be
A=\dfrac{4}{3\gamma},
\label{A}
\ee
and the various coefficients in the formal series (\ref{seriesgab}) are inhomogeneous functions of the space coordinates.
Also, from Eqs. (\ref{cGseries}), (\ref{abeh}), and (\ref{A}), we have,
\be
c=c_{0}t^{An/2}, \quad G=G_{0}t^{Am/2}, \quad n,m\in \mathbb {R}.
\label{cG}
\ee

Using the expansion (\ref{seriesgab}), the \emph{inverse} metric tensor components read
\be
\gamma ^{\alpha \beta } = a^{\alpha \beta }t^{-A}-b^{\alpha \beta}t^{B-2A}+\cdots,
\label{inverse seriesgab}
\ee

where $\gamma _{\alpha \beta }\gamma ^{\beta \gamma }=\delta _{\alpha}^{\gamma }$, and  $a_{\alpha \beta }a^{\beta \gamma }=\delta_{\alpha }^{\gamma }$.
The indices of $b_{\alpha \beta }$ and all other matrices in the expansion of the spatial metric are raised by $a^{\alpha \beta}$.
In addition, using the series representations for $\gamma_{\alpha \beta }$ and its reciprocal, we can write down the series expressions for the extrinsic curvature and the various series related to it and appear in the system of equations (\ref{field00})-(\ref{fieldab}).
This leads to
\bq
K_{\alpha \beta } &=& \partial_t \gamma _{\alpha \beta } = Aa_{\alpha \beta }t^{A-1} + Bb_{\alpha \beta }t^{B-1} + \cdots, \label{extrin_series}\\
K^{\beta }_{\alpha } &=& \gamma^ {\beta \mu} K_{\mu \alpha} = A\delta^{\beta }_{\alpha }t^{-1} + (B-A)b^{\beta }_{\alpha }t^{B-A-1} + \cdots, \label{kab} \\
K &=& K^\alpha_\alpha = 3At^{-1} + (B-A)bt^{B-A-1} + \cdots.
\label{k}
\eq
Then the time derivatives are given by,
\bq
\partial_t K^{\beta }_{\alpha } &=& -A\delta^{\beta }_{\alpha }t^{-2} + (B-A-1)(B-A)b^{\beta }_{\alpha }t^{B-A-2} + \cdots, \\
\label{dermixK}
\partial_t K &=& -3At^{-2} + (B-A-1)(B-A)bt^{B-A-2} + \cdots.
\label{derK}
\eq

Below, we shall also present the expressions of the products that will appear in our system of equations, namely the expansions,
\bq 
K_{\alpha }^{\beta }K_{\beta }^{\alpha } &=& 3A^{2}t^{-2} + 2A(B-A)bt^{B-A-2} + \cdots, \\
\label{product mixed K}
KK^{\beta }_{\alpha } &=& 3A^{2}\delta^{\beta }_{\alpha }t^{-2} + A(B-A)(3b^{\beta }_{\alpha } + b\delta^{\beta }_{\alpha })t^{B-A-2} + \cdots.
\label{product K mixed K}
\eq 

\subsection{Ricci tensor and perfect fluid}
The previous expressions (\ref{extrin_series})-(\ref{product K mixed K}) will be substituted into the components of the Ricci tensor $R_{ij}$ and its trace $R$, in order to use them in the Einstein field equations (\ref{field00})-(\ref{fieldab}).
The formulas for the mixed Ricci tensor are given by (see \cite{13}),

\begin{eqnarray}
R_{0}^{0} &=& -\frac{1}{2c^2}\partial_t K - \frac{1}{4c^2}K^\beta_\alpha K^\alpha_\beta  ,\label{seriesR00} \\
R_{\alpha }^{0} &=& \frac{1}{2c^2}(\nabla_\beta K^\beta_\alpha - \nabla_\alpha K),  \label{seriesR0a}\\
R^{\beta }_{\alpha } &=& -P^\beta_\alpha - \frac{1}{2c^2}\partial_t K^\beta_\alpha - \frac{1}{4c^2}KK^\beta_\alpha,
\label{rab}
\end{eqnarray}
and the trace is,
\be
R = -P -\frac{1}{c^2}\partial_t K -\frac{1}{4c^2}K^{2} -\frac{1}{4c^2}K^\beta_\alpha K^\alpha_\beta.
\label{R}
\ee
In the Appendix, we give the series expansions related to the 3-dimensional Ricci tensor $P_{\alpha \beta}$ and its trace $P$.\\
Now, taking into account the relations (\ref{eqstate}) and (\ref{velocities}), we can write down the expressions for the components of the mixed stress-energy tensor given by (\ref{stress-energy tensor}).
Considering higher-order corrections, we get,
\begin{eqnarray}
T_{0}^{0} &=&(\rho + \frac{p}{c^2})u^0u_{0}-p\sim\rho c^2 ,
\label{T00} \\
T_{\alpha }^{0} &=&(\rho + \frac{p}{c^2})u^{0}u_{\alpha }\sim \gamma \rho u_{\alpha },
\label{T0a} \\
T^{\beta }_{\alpha } &=&(\rho + \frac{p}{c^2})u^{\beta }u_{\alpha }-p\delta ^{\beta}_{\alpha }\sim (1-\gamma)\rho c^{2} \delta ^{\beta }_{\alpha },
\label{Tab}
\end{eqnarray}
and also,
\be
T = (4-3\gamma)\rho c^{2}.
\label{T}
\ee

\subsection{Field equations and final balance}
Starting from the Einstein field equations (\ref{field00})-(\ref{fieldab}), and using the various relations of the terms that appear in them, we can find expressions associated with the energy density $\rho c^2$, the three-vector $u_{\alpha}$, as well as the connections between the spatial tensors $a_{\alpha \beta}$ and $b_{\alpha \beta}$.
We are faced with the problem of finding the leading orders of the various terms in the split field equations given before.
After imposing the series given in (\ref{seriesgab}) into equation (\ref{field00}) for $\gamma \neq 2/3 \quad (A\neq2)$, we obtain the following relation,
\begin{eqnarray}
\rho &=&  \dfrac{1}{4\pi (3\gamma - 2)G} \left(-\frac{1}{2}\partial_t K - \frac{1}{4}K^\beta_\alpha K^\alpha_\beta \right)
\nonumber \\
&=&\dfrac{1}{16\pi (3\gamma - 2)G_{0}}\bigg[ 3A(2-A)t^{-2-\frac{Am}{2}} - 2(B-A)(B-1)bt^{B-A-2-\frac{Am}{2}}  \nonumber \\
&+&  \cdots \bigg],
\label{seriesdensity}
\end{eqnarray}
which is not the final expression for $\rho$, since we are going to associate the parameter $B$ with $A$, as well as the spatial tensor $b_{\alpha \beta}$ with $a_{\alpha \beta}$.
This will be possible using the `evolution' equations of the Einstein field equations.
In fact, from the $\binom{\beta}{\alpha}$ components of the Ricci tensor, given by (\ref{fieldab}), we find,
\be
-c^{2}P^{\beta}_{\alpha} - \frac{1}{2}\partial_t K^{\beta}_{\alpha} - \frac{1}{4}KK^{\beta}_{\alpha} = 4\pi G (\gamma - 2)\rho \delta^{\beta}_{\alpha}.
\label{seriespressure}
\ee
Apart from $\gamma \neq 2/3$, if we also require $\gamma \neq 2$ and use (\ref{seriesdensity}), we get,
\bq
&-&{c_{0}}^{2}(P^{\beta}_{\alpha})_{-A}t^{A(n-1)} + \frac{1}{4}A(2-3A)\delta^{\beta}_{\alpha} t^{-2} - \frac{1}{4}(B-A)\big[ (A+2B-2)b^{\beta}_{\alpha} +A\delta^{\beta}_{\alpha} \big]t^{B-A-2} \nonumber\\
&=& \dfrac{\gamma - 2}{4(3\gamma - 2)}\Big[ 3A(2-A)t^{-2} - 2(B-A)(B-1)bt^{B-A-2} \Big]\delta^{\beta}_{\alpha} + \cdots.
\label{balanceab}
\eq
But, $\gamma = \frac{4}{3A}$, so the terms of order $t^{-2}$ cancel.
Also, in order to keep the balance in the series, we must have,
\be
B = 2 + An = 2 + \frac{4n}{3\gamma},
\label{B}
\ee
and then the remaining terms proportional to $t^{A(n-1)}$ give,
\be
b^{\beta}_{\alpha} = -\dfrac{4{c_{0}}^{2}(P^{\beta}_{\alpha})_{-A}}{(2-A+An)(2+A+2An)} + \Big[ \dfrac{2(2-3A)(1+An)}{3(2-A)} - A \Big]\dfrac{b\delta^{\beta}_{\alpha}}{A+2+2An}.
\label{bbetaalpha1}
\ee
Because of the existence of both $b^{\beta}_{\alpha}$ and $b$ in (\ref{bbetaalpha1}), we can use trace in order to eliminate $b$.
Consequently, we find,
\be
b = -\dfrac{{c_{0}}^{2}(A-2)}{A(2-A+An)(3-A+An)}P_{-A}.
\label{b}
\ee
Therefore, substituting (\ref{b}) into (\ref{bbetaalpha1}) and using (\ref{A}), we finally find that,
\be
b^{\beta}_{\alpha} = -\dfrac{9{\gamma}^{2}{c_{0}}^{2}}{4(3\gamma -2+2n)} \Big[ \dfrac{4(P^{\beta}_{\alpha})_{-A}}{3\gamma +2+8n} + \dfrac{(\gamma -2)(3\gamma +4n) - 2(3\gamma -2)}{(3\gamma +2+4n)(9\gamma -4+4n)}P_{-A}\delta^{\beta}_{\alpha} \Big].
\label{bbetaalpha2}
\ee
In this last relation, we see that the components of the spatial tensor $b_{\alpha \beta}$ are completely determined by $a_{\alpha \beta}$, because both $(P^{\beta}_{\alpha})_{-A}$ and $P_{-A}$ depend only on $a_{\alpha \beta}$ (see Appendix).
Notice, also, that out of the two parameters of the function $G(t)$, neither $G_0$ nor $m$ is present in the relation (\ref{bbetaalpha2}).
As mentioned above, it is possible to find a simplified expression for $\rho$.
We can use (\ref{b}) together with (\ref{B}) to get the final expression resulting from the $\binom{0}{0}$ component of the Ricci tensor, namely that,
\be
\rho = \dfrac{1}{6\pi {\gamma}^{2}G_{0}}\bigg[ t^{-2-\frac{2m}{3\gamma}} + \dfrac{3{\gamma}^{2}(3\gamma + 4n){c_0}^{2}P_{-A}}{8(9\gamma - 4 + 4n)}t^{\frac{4}{3\gamma}(n-1-\frac{m}{2})} +  \cdots \bigg].
\label{seriesdensity2}
\ee
Here, only the matrix $a_{\alpha \beta}$ appears together with various parameters.
Notice that the leading order of (\ref{seriesdensity2}) does not depend on the time-varying $c$, that is the result would be qualitatively the same in a constant $c$-varying $G$ model.
Furthermore, notice that the result for $\rho$ is consistent with that obtained in \cite{8}.
In order to find the expression for the velocities $u_\alpha$, we need to compute the difference $\nabla_{\beta}b^{\beta}_{\alpha} - \nabla_{\alpha}b$ that will appear in the $\binom{0}{\alpha}$ constraints, thus from (\ref{bbetaalpha2}), we find,
\be
\nabla_{\beta}b^{\beta}_{\alpha} = -\dfrac{9{\gamma}^{2}{c_{0}}^{2}}{4(3\gamma -2+2n)} \Big[ \dfrac{4\nabla_{\beta}(P^{\beta}_{\alpha})_{-A}}{3\gamma +2+8n} + \dfrac{(\gamma -2)(3\gamma +4n) - 2(3\gamma -2)}{(3\gamma +2+4n)(9\gamma -4+4n)}\nabla_{\alpha}P_{-A} \Big],
\label{nablabetaalpha}
\ee
but, in view of the identity \cite{13},
\be
\nabla_{\beta}P^{\beta}_{\alpha} = \dfrac{1}{2}\nabla_{\alpha}P,
\label{identityP}
\ee
we find that,
\be
\nabla_{\beta}b^{\beta}_{\alpha} = -\dfrac{9{\gamma}^{2}{c_{0}}^{2}}{4(3\gamma -2+2n)} \Big[ \dfrac{2}{3\gamma +2+8n} + \dfrac{(\gamma -2)(3\gamma +4n) - 2(3\gamma -2)}{(3\gamma +2+4n)(9\gamma -4+4n)} \Big] \nabla_{\alpha}P_{-A}.
\label{nablabetaalpha2}
\ee
Therefore, taking into account (\ref{b}) and (\ref{A}), we conclude that,
\begin{eqnarray}
\nabla_{\beta}b^{\beta}_{\alpha} - \nabla_{\alpha}b &=& -\dfrac{9{\gamma}^{2}{c_{0}}^{2}}{4(3\gamma -2+2n)} \Big[ \dfrac{2}{3\gamma +2+8n} + \dfrac{(\gamma -2)(3\gamma +4n) - 2(3\gamma -2)}{(3\gamma +2+4n)(9\gamma -4+4n)} \nonumber \\
&+& \dfrac{3\gamma - 2}{9\gamma - 4 + 4n} \Big] \nabla_{\alpha}P_{-A}.
\label{difference}
\end{eqnarray}
Then, using the $\binom{0}{\alpha}$ components of the Ricci tensor, given by (\ref{field0a}), we get,
\be
\nabla_{\beta}K^{\beta}_{\alpha} - \nabla_{\alpha}K = \dfrac{16 \pi G}{c^2}\gamma \rho u_{\alpha}.
\label{seriesvel}
\ee
From the expressions (\ref{cG}) for $c$ and $G$, and the previous relations (\ref{seriesdensity2}) and (\ref{difference}), we find that the velocities satisfy,
\begin{eqnarray}
u_{\alpha} &=& -\dfrac{9{\gamma}^{2}{c_{0}}^{4}}{16} \Big[ \dfrac{2}{3\gamma +2+8n} + \dfrac{(\gamma -2)(3\gamma +4n) - 2(3\gamma -2)}{(3\gamma +2+4n)(9\gamma -4+4n)} \nonumber \\
&+& \dfrac{3\gamma - 2}{9\gamma - 4 + 4n} \Big] \nabla_{\alpha}P_{-A}\;
t^{3 + 4\frac{n-1}{3\gamma}}.
\label{vel}
\end{eqnarray}
Notice that, in (\ref{vel}), $u_{\alpha}$ depends only on the spatial tensor $a_{\alpha \beta}$ plus the parameters $\gamma$, $n$ and $c_0$, but does not depend at all on the parameters $G_{0}$ and $m$ that are associated with the gravitational `constant' $G$.

\subsection{Parameter domains and series consistency}
To ensure a mathematically robust and physically consistent formal asymptotic expansion near the initial cosmic singularity, the parameter space $(\gamma, n, m)$ must satisfy specific conditions:

Firstly, regarding the ordering of the metric series ($A < B$): Since $A = \frac{4}{3\gamma}$ and the evolution equations dictate the balance $B = 2 + An = 2 + \frac{4n}{3\gamma}$, the condition for $b_{\alpha\beta}$ to act as a genuine subleading perturbation ($B > A$) requires:
\begin{equation}
2 + \frac{4n}{3\gamma} > \frac{4}{3\gamma} \implies n > 1 - \frac{3}{2}\gamma.
\end{equation}
For a radiation-dominated universe ($\gamma = 4/3$), this reduces elegantly to $n > -1$.

Secondly, regarding the absence of resonances: From the explicit algebraic coefficients for the subleading tensor $b^\beta_\alpha$ and velocity $u_\alpha$, we must strictly avoid poles in the parameter coefficients, yielding the resonance conditions:
\begin{equation}
\gamma \neq \frac{2}{3}, \quad 3\gamma - 2 + 2n \neq 0, \quad 9\gamma - 4 + 4n \neq 0.
\end{equation}

Finally, examining the behavior for large negative $n$ ($n \le 1 - \frac{3}{2}\gamma$), the structural ordering breaks down ($B \le A$). Physically, if the speed of light drops precipitously as the universe expands (or grows extremely fast going backward toward the singularity), the spatial curvature and VSL coupling terms grow faster than the kinetic driving terms. In this regime, the standard quasi-isotropic ansatz fails, and the system shifts to a completely different asymptotic balance where VSL modifications dominate the leading-order singularity dynamics.

\subsection{Final counting at the initial singularity}
In our system of equations (\ref{field00})-(\ref{fieldab}), there are $16$ initial functions involved.
These functions are the twelve components of the spatial matrices $a_{\alpha \beta}$ and $b_{\alpha \beta}$ that appear in the series expansion of the metric $\gamma_{\alpha \beta}$, the three velocities $u_{\alpha}$, and the function $\rho$.
Regarding the time-varying parameters $c(t)$ and $G(t)$, although they introduce degrees of freedom, in our model they are constrained by (\ref{cG}).
Therefore, they do not contribute as independent functions in our counting for the Cauchy problem.
In other words, since $c$ and $G$ are assumed to be spatially homogeneous, depending only on the time coordinate restricted by (\ref{cG}), they do not finally provide additional degrees of freedom.
Even if we consider $c$ and $G$ as potential fields, which would increase the number of arbitrary functions required for a general solution from $8$ to $10$, the constraints (\ref{cG}) would reduce this number back to $8$.
From (\ref{bbetaalpha2}), all six components of $b_{\alpha \beta}$ are fully determined by $a_{\alpha \beta}$, while the six components of $a_{\alpha \beta}$ remain arbitrary.
Next, taking into account (\ref{seriesdensity2}) and (\ref{vel}), the initial functions $\rho$ and $u_{\alpha}$ are also fully determined by the spatial matrix $a_{\alpha \beta}$.
Although the various parameters $c_{0}$, $G_{0}$, $m$, and $n$ are present in the results, they are constants of the model and therefore do not count as additional arbitrary functions.
Thus, there remain only $6$ arbitrary functions. However, the space coordinates still permit arbitrary transformations that do not depend on time.
By subtracting these $3$ spatial coordinate covariances, which may be used to bring $a_{\alpha \beta}$ to diagonal form, we finally find that the given solution involves $3$ independent functions of the spatial coordinates.
Regarding the fourth coordinate covariance that is related to the time coordinate, and taking into consideration (\ref{seriesgab}) and the fact that $A \neq 0$, we conclude that the time in the metric form is completely determined by the condition $t=0$ at the singularity.
Therefore, we do not further decrease the number of arbitrary functions.
The resulting count of $3$ functions falls short of the number required for the general solution.
As we have shown, the general solution of the system must contain $8$ free functions.

\section{Discussions and Conclusions}

In this paper, we have explored the behavior of the asymptotic solutions in varying speed of light cosmologies within the context of the Landau-Lifshitz quasi-isotropic method. We have shown that perturbations of the isotropic solution near the initial singularity ($a \sim t^{2/3\gamma}$), in the framework of VSL theory, do not lead to a generic solution as indicated by the number of arbitrary functions found in the final solution. In particular, the coupling of the time-dependent $c(t)$ and/or $G(t)$ with the analytic expansion of the spatial metric results in fewer independent arbitrary functions than those expected in a general VSL solution.

Specifically, our order-by-order function counting demonstrates that the quasi-isotropic VSL solution yields only 3 independent arbitrary functions of the spatial coordinates, falling short of the 8 free functions required to qualify as a truly general cosmological solution. This mathematically proves that the quasi-isotropic ansatz remains structurally non-generic under varying fundamental `constants', directly mirroring the behavior found in standard General Relativity. Our core conclusion is completely robust and insensitive to the specific power-law forms chosen for $c(t)$ and $G(t)$, since these introduce only discrete constant parameters and carry zero functional degrees of freedom over the spatial coordinates.

Furthermore, it is crucial to contextualize these findings within the broader framework of the classic Belinskii-Khalatnikov-Lifshitz (BKL) chaotic dynamics near a spacelike singularity. While our quasi-isotropic analysis assumes a smooth, symmetric approach to the singularity, a complete understanding of early-universe VSL dynamics requires exploring the full anisotropic, chaotic regime associated with the mixmaster behavior. Interestingly, the introduction of a dynamically varying speed of light could qualitatively alter this picture. As $c(t) \to \infty$ near the initial singularity, the comoving causal horizon expands dramatically, enhancing causal communication across spatial points and allowing spatial gradients to remain highly active. This horizon growth acts as a potential mechanism that could suppress the chaotic BKL mixmaster transitions, potentially driving the universe toward isotropy without fine-tuned initial data. A rigorous mathematical formulation of the full anisotropic BKL equations and the corresponding cosmological billiard dynamics under a varying-$c$ framework represents a major open challenge, and the non-generic tensorial expansions presented here serve as a necessary stepping stone toward that direction.

Additionally, regarding the excluded cases $\gamma = 2/3$ and $\gamma = 2$, these thresholds are of particular physical and mathematical interest. Preliminary analysis suggests that these specific values may alter the asymptotic behavior of the cosmological models, potentially necessitating the inclusion of logarithmic terms or resonant structures to achieve a consistent mathematical balance. A full analysis of these special cases is the subject of a subsequent, specialized study.

Finally, as a cross-verification, transitioning to a unit-lapse cosmic time gauge by defining $d\tau = c(t) dt$ provides an elegant alternative mathematical representation that significantly streamlines the gravitational algebra. In this unit-lapse frame, the lapse is identically unity, naturally eliminating explicit time derivatives of $c$ from the Ricci tensor components. The profound physical effects of VSL are then cleanly isolated within the effective time-dependent coupling constant $\kappa(\tau) = \frac{8\pi G(\tau)}{c(\tau)^4}$ and the modified local non-conservation laws, confirming that the core physics of these VSL models is independent of the coordinate representation.

This work bridges the gap between the method of asymptotic splittings used in \cite{8} and the asymptotic tensorial method introduced in \cite{13} and \cite{14} for VSL models. In fact, while a Fuchsian series expansion corresponding to admissible asymptotic isotropic solutions was found in \cite{8}, the method proposed in this paper provides asymptotic anisotropic tensorial series solutions.

\appendix
\section{Spatial Ricci expansion}

As discussed in Section 2, the setting of $\Gamma^0_{00}=0$ is rigorously justified either by adopting the structural prescription framework where $c(t)$ is treated algebraically during connection computation, or by transforming to a unit-lapse cosmic time gauge via $d\tau=c(t)dt$.

The purpose of this Appendix is to prove that the form of the terms $(P^{\beta}_{\alpha})_{-A}$ and $P_{-A}$ depends only on the spatial matrix $a_{\alpha \beta}$.
We also give details for the next terms of the 3-dimensional Ricci tensor $P_{\alpha \beta}$ and its trace $P$.
To this end, we consider a synchronous reference system, whose 4-dimensional Christoffel symbols are given by,
\begin{equation}
\Gamma^i_{jk}=\frac{1}{2}g^{il}(\partial_k g_{lj} + \partial_j g_{lk} - \partial_l g_{jk}).
\label{Chrostffel}
\end{equation}
Therefore, the corresponding components of $\Gamma^i_{jk}$ are the following,
\begin{eqnarray}
\Gamma^0_{00} &=& \Gamma^\alpha_{00} = \Gamma^0_{0 \alpha} = 0, \\
\Gamma^0_{\alpha \beta} &=& \frac{1}{2c^2}K_{\alpha \beta} = \frac{1}{2{c_{0}}^2}Aa_{\alpha \beta}\;t^{A-1-An} + \frac{1}{2{c_{0}}^2}Bb_{\alpha \beta}\;t^{B-1-An} + \cdots, \\
\Gamma^\beta_{0 \alpha} &=& \frac{1}{2}K^\beta_\alpha = \frac{1}{2}A\delta^{\beta }_{\alpha }\;t^{-1} + \frac{1}{2}(B-A)b^\beta_\alpha\;t^{B-A-1} + \cdots \\
\Gamma^\alpha_{\beta \gamma} &=& \frac{1}{2}\gamma^{\alpha \delta}(\partial_\gamma \gamma_{\delta \beta} + \partial_\beta \gamma_{\delta \gamma} - \partial_\delta \gamma_{\beta \gamma}),
\end{eqnarray}
where the last ones are the 3-dimensional Christoffel symbols formed using the metric $\gamma_{\alpha\beta}$.
The three-dimensional Ricci tensor $P_{\alpha \beta}$ associated with $\gamma_{\alpha \beta}$ is then given by,
\begin{equation}
P_{\alpha \beta} = \partial_\mu \Gamma^\mu_{\alpha \beta} - \partial_\beta \Gamma^\mu_{\alpha \mu} +
\Gamma^\mu_{\alpha \beta}\Gamma^\epsilon_{\mu \epsilon} - \Gamma^\mu_{\alpha \epsilon}\Gamma^\epsilon_{\beta \mu}.
\label{eqP_ab}
\end{equation}

It is important to start with the asymptotic expansions of the spatial Ricci tensor $P_{\alpha \beta}$ and its trace near the singularity.
This is done below in several steps. First, we introduce the symbols,
\begin{eqnarray}
A_{\alpha \beta \epsilon} &=& \partial_\beta a_{\alpha \epsilon} + \partial_\alpha a_{\beta \epsilon} - \partial_\epsilon a_{\alpha \beta}, \\
B_{\alpha \beta \epsilon} &=& \partial_\beta b_{\alpha \epsilon} + \partial_\alpha b_{\beta \epsilon} - \partial_\epsilon b_{\alpha \beta},
\end{eqnarray}
and using the basic metric expansion (\ref{seriesgab}) around the singularity, we find that,
\begin{eqnarray}
\Gamma^\alpha_{\beta\gamma} &=&
\frac{1}{2}a^{\alpha \delta}A_{\beta\gamma\delta} + \frac{1}{2}(a^{\alpha \delta} B_{\beta\gamma\delta} - b^{\alpha \delta} A_{\beta\gamma\delta})\;t^{B-A} + \cdots \nonumber \\
&=& (\Gamma^\alpha_{\beta\gamma})_{0} + (\Gamma^\alpha_{\beta\gamma})_{B-A}\;t^{B-A} + \cdots,
\label{seriesG}
\end{eqnarray}
with the coefficients $(\Gamma^\alpha_{\beta\gamma})_{0}$ and $(\Gamma^\alpha_{\beta\gamma})_{B-A}$ clearly defined.
Then Eq. (\ref{eqP_ab}) implies,
\begin{equation}
P_{\alpha \beta} = (P_{\alpha \beta})_{0} + (P_{\alpha \beta})_{B-A}\;t^{B-A} + \cdots,
\label{seriesPab}
\end{equation}
where the corresponding coefficients of the first two orders for the spatial Ricci 3-curvature are expressed using the $(\Gamma^\mu_{\alpha \beta})_i$'s as follows,
\begin{eqnarray}
(P_{\alpha \beta})_0 &=& \partial_\mu (\Gamma^\mu_{\alpha \beta})_0 -
\partial_\beta (\Gamma^\mu_{\alpha \mu})_0 +
(\Gamma^\mu_{\alpha \beta})_0 (\Gamma^\epsilon_{\mu \epsilon})_0 -
(\Gamma^\mu_{\alpha \epsilon})_0 (\Gamma^\epsilon_{\beta \mu})_0, \\
(P_{\alpha \beta})_{B-A} &=& \partial_\mu (\Gamma^\mu_{\alpha \beta})_{B-A} - \partial_\beta (\Gamma^\mu_{\alpha \mu})_{B-A} +
(\Gamma^\mu_{\alpha \beta})_0 (\Gamma^\epsilon_{\mu \epsilon})_{B-A} \nonumber \\
&+& (\Gamma^\epsilon_{\mu \epsilon})_0 (\Gamma^\mu_{\alpha \beta})_{B-A}
-(\Gamma^\mu_{\alpha \epsilon})_0 (\Gamma^\epsilon_{\beta \mu})_{B-A} -
(\Gamma^\epsilon_{\beta \mu})_0 (\Gamma^\mu_{\alpha \epsilon})_{B-A}.
\end{eqnarray}
Then the mixed components of the Ricci 3-curvature are given by,
\begin{eqnarray}
P^\beta_\alpha &=& \gamma^{\beta \mu} P_{\mu \alpha} \nonumber \\
&=& a^{\beta \mu}(P_{\mu \alpha})_{0}\;t^{-A} + [a^{\beta \mu}(P_{\mu \alpha})_{B-A} - b^{\beta \mu}(P_{\mu \alpha})_0]\;t^{B-2A} + \cdots \nonumber \\
&=& (P^\beta_\alpha)_{-A}\;t^{-A} + (P^\beta_\alpha)_{B-2A}\;t^{B-2A} + \cdots.
\label{seriesPabmixed}
\end{eqnarray}
This result implies that the trace expansion for the $P=\textrm{Tr}P^\alpha_\beta$ is given by,
\begin{eqnarray}
P &=&\delta^\alpha_\beta P^\beta_\alpha \nonumber \\
&=& a^{\alpha \mu}(P_{\mu \alpha})_{0}\;t^{-A} + [a^{\alpha \mu}(P_{\mu \alpha})_{B-A} - b^{\alpha \mu}(P_{\mu \alpha})_0]\;t^{B-2A} + \cdots \nonumber \\
&=& P_{-A}t^{-A} + P_{B-2A} t^{B-2A} + \cdots.
\label{seriesP}
\end{eqnarray}
Thus, the terms that we are mainly interested in, namely the terms $(P^\beta_\alpha)_{-A}$ and $P_{-A}$ are then given explicitly by the expansions,
\begin{eqnarray}
(P^\beta_\alpha)_{-A} &=& a^{\beta \gamma}(P_{\gamma \alpha})_{0} \nonumber \\
&=& \frac{1}{2}a^{\beta \gamma} \Big[ \partial_{\mu}(a^{\mu \delta}A_{\alpha \gamma \delta}) -  \partial_{\alpha}(a^{\mu \delta}A_{\gamma \mu \delta}) + \frac{1}{2}a^{\mu \delta}a^{\epsilon \zeta}(A_{\alpha \gamma \delta}A_{\mu \epsilon \zeta} - A_{\gamma \epsilon \delta}A_{\alpha \mu \zeta}) \Big] \nonumber \\
&=& \frac{1}{2}a^{\beta \gamma} \Big\{ \partial_{\mu} \Big[ a^{\mu \delta}(\partial_{\gamma}a_{\alpha \delta} + \partial_{\alpha}a_{\gamma \delta} - \partial_{\delta}a_{\alpha \gamma}) \Big] - \partial_{\alpha} \Big[ a^{\mu \delta}(\partial_{\mu}a_{\gamma \delta} + \partial_{\gamma}a_{\mu \delta} - \partial_{\delta}a_{\gamma \mu}) \Big] \nonumber \\
&+& \frac{1}{2}a^{\mu \delta}a^{\epsilon \zeta} \Big[ (\partial_{\gamma}a_{\alpha \delta} + \partial_{\alpha}a_{\gamma \delta} - \partial_{\delta}a_{\alpha \gamma})(\partial_{\epsilon}a_{\mu \zeta} + \partial_{\mu}a_{\epsilon \zeta} - \partial_{\zeta}a_{\mu \epsilon}) \nonumber \\
&-& (\partial_{\epsilon}a_{\gamma \delta} + \partial_{\gamma}a_{\epsilon \delta} - \partial_{\delta}a_{\gamma \epsilon})(\partial_{\mu}a_{\alpha \zeta} + \partial_{\alpha}a_{\mu \zeta} - \partial_{\zeta}a_{\alpha \mu}) \Big] \Big\},
\label{eqPba-A}
\end{eqnarray}
and,
\begin{eqnarray}
P_{-A} &=& a^{\alpha \beta}(P_{\alpha \beta})_0 \nonumber \\
&=& \frac{1}{2}a^{\alpha \beta}[\partial_\mu (a^{\mu \delta}A_{\alpha \beta \delta}) -
\partial_\beta (a^{\mu \delta}A_{\alpha \mu \delta}) + \frac{1}{2}a^{\mu \delta}a^{\epsilon \zeta}(A_{\alpha \beta \delta}A_{\mu \epsilon \zeta} - A_{\beta \epsilon \delta}A_{\alpha \mu \zeta})] \nonumber \\
&=& \frac{1}{2}a^{\alpha \beta} \Big\{ \partial_\mu [a^{\mu \delta} (\partial_\beta a_{\alpha \delta} + \partial_\alpha a_{\beta \delta} - \partial_\delta a_{\alpha \beta})]  - \partial_\beta [a^{\mu \delta} (\partial_\mu a_{\alpha \delta} + \partial_\alpha a_{\mu \delta} - \partial_\delta a_{\alpha \mu})] \nonumber \\
&+& \frac{1}{2}a^{\mu \delta}a^{\epsilon \zeta} \Big[ (\partial_{\beta}a_{\alpha \delta} + \partial_{\alpha}a_{\beta \delta} - \partial_{\delta}a_{\alpha \beta})(\partial_{\epsilon}a_{\mu \zeta} + \partial_{\mu}a_{\epsilon \zeta} - \partial_{\zeta}a_{\mu \epsilon}) \nonumber \\
&-& (\partial_{\epsilon}a_{\beta \delta} + \partial_{\beta}a_{\epsilon \delta} - \partial_{\delta}a_{\beta \epsilon})(\partial_{\mu}a_{\alpha \zeta} + \partial_{\alpha}a_{\mu \zeta} - \partial_{\zeta}a_{\alpha \mu}) \Big] \Big\}.
\label{eqP0}
\end{eqnarray}

It is clear from (\ref{eqPba-A}) and (\ref{eqP0}) that both $(P^\beta_\alpha)_{-A}$ and $P_{-A}$ consist only of combinations of the spatial matrix $a_{\alpha \beta}$ and its spatial derivatives.
Furthermore, using (\ref{seriesPabmixed}) and (\ref{seriesP}), it is obvious that the terms proportional to $t^{B-2A}$ in the series of  $P^{\beta}_{\alpha}$ and $P$ depend only on the matrices $a_{\alpha \beta}$ and $b_{\alpha \beta}$ and their spatial derivatives.

\end{document}